\documentclass[prd,twocolumn,preprint numbers,aps,superscriptaddress,nofootinbib,showpacs]{revtex4-1}
 \usepackage{amsmath}
\usepackage{amssymb}
\usepackage{graphicx}
\usepackage{soul,color}
\usepackage{bm}
\usepackage{times}
\usepackage{hyperref}
\hypersetup{
  colorlinks=true,
  citecolor=blue,
  linkcolor=blue,
  urlcolor=blue}

\usepackage{epsfig}% Include figure files
\usepackage{dcolumn}% Align table columns on decimal point
\usepackage{float}

\usepackage{epsfig}% Include figure files
\usepackage{dcolumn}% Align table columns on decimal point
\def\be{\begin{equation}}
\def\ee{\end{equation}}
\def\bea{\begin{eqnarray}}
\def\eea{\end{eqnarray}}
\def\gsim{\ \rlap{\raise 2pt\hbox{$>$}}{\lower 2pt \hbox{$\sim$}}\ }
\def\lsim{\ \rlap{\raise 2pt\hbox{$<$}}{\lower 2pt \hbox{$\sim$}}\ }
\def\dslash{\kern-4pt \not{\hbox{\kern-2pt $\partial$}}}
\def\pslash{\not{\hbox{\kern-2pt p}}}

\begin{document}

\setstcolor{red}

\DeclareGraphicsExtensions{.eps,.ps}

\title{Texture zeros of low-energy Majorana neutrino mass matrix in 3+1 scheme}

\author{Debasish Borah}
\email{dborah@iitg.ernet.in}
\affiliation{Department of Physics, Indian Institute of Technology Guwahati, 
Assam-781039, India}

\author{Monojit Ghosh}
\email{mghosh@phys.se.tmu.ac.jp}
\affiliation{Department of Physics, Tokyo Metropolitan University, Hachioji, 
Tokyo 192-0397, Japan}

\author{Shivani Gupta}
\email{shivani.gupta@adelaide.edu.au}
\affiliation{
Center of Excellence in Particle Physics (CoEPP), University of Adelaide, 
Adelaide SA 5005, Australia}

\author{Sushant K. Raut}
\email{sushant@ibs.re.kr}
\affiliation{Center for Theoretical Physics of the Universe,
Institute for Basic Science (IBS), Daejeon, 34051, Korea}

%\maketitle
\begin{abstract}
In this work we revisit the zero textures in low energy 
Majorana neutrino mass matrix when the active neutrino sector
is extended by a light sterile neutrino in the eV scale i.e., the 3+1 scheme. In 
3+1 scenario, the low energy neutrino mass matrix ($m_\nu$) has ten independent 
elements. Thus in principle one can have 
minimum one-zero texture to maximum ten-zero texture. We summarize the previous 
results of one, two, three and four-zero textures which already exist in the 
literature and present our 
new results on five-zero textures. In our analysis we find that among six possible 
five-zero textures, only one is allowed by the present data. We 
discuss possible theoretical model which can explain the origin of the 
allowed five-zero texture and discuss other possible implications of such a 
scenario. Our results also concludes that in 3+1 scheme, one can not have more 
than five-zeros in $m_\nu$.

\end{abstract}

\preprint{ADP-17 - 24 / T1030}
\preprint{CTPU-17-19}
\maketitle

\section{Introduction}
\label{sec1}

The possibility of light sterile neutrinos with mass at the eV scale have 
gathered serious attention in the last two decades following the neutrino 
anomalies reported by some experiments 
which could be explained by incorporating additional light neutrinos to which 
the active neutrinos can oscillate into. For a review, one may refer to 
Ref.~\cite{Abazajian:2012ys}. 
The first such anomaly was reported by the Liquid Scintillator Neutrino Detector 
(LSND) experiment in their anti-neutrino flux measurements~\cite{Athanassopoulos:1996jb,Aguilar:2001ty}. 
The LSND experiment searched for $\bar{\nu}_{\mu} \rightarrow \bar{\nu}_e$ 
oscillations in the appearance mode and reported an excess of $\bar{\nu}_e$ 
interactions that could be explained 
by incorporating at least one additional light neutrino with mass in the eV 
range. This result was supported by the subsequent measurements at the MiniBooNE 
experiment~\cite{Aguilar-Arevalo:2013pmq}. 
Similar anomalies have also been observed at reactor neutrino
experiments~\cite{Mention:2011rk} as well as gallium solar neutrino 
experiments~\cite{Acero:2007su,Giunti:2010zu}. 
Since the precision measurements at the LEP experiment do not allow additional 
light neutrinos coupling to the standard model (SM) gauge bosons~\cite{Agashe:2014kda}, 
such additional light neutrinos are 
called sterile neutrinos. These anomalies require the presence of a light 
sterile neutrino at eV scale having non-trivial mixing with the active neutrinos 
as presented 
in the global fit studies~\cite{Kopp:2013vaa,Giunti:2013aea,Gariazzo:2015rra}. 

Apart from these reactor and accelerator based experiments, there were initial 
hints from cosmology as well, suggesting the presence of one additional light 
neutrino. 
For example, the nine year Wilkinson Microwave Anisotropy Probe (WMAP) data 
suggested the total number of relativistic degrees of freedom to be 
$N_{\text{eff}} = 3.84 \pm 0.40$~\cite{Hinshaw:2012aka}. 
Since the standard value is $N_{\text{eff}} = 3.046$, the WMAP data could 
accommodate one additional light species. 
Such cosmology experiments can constrain the number of such relativistic degrees 
of freedom as they affect the big bang nucleosynthesis (BBN) predictions as well 
as cause changes in the cosmic 
microwave background (CMB) spectrum, which are very accurately measured. 
Contrary to the WMAP findings, the more recent Planck experiment puts $95\%$ 
limit on the effective number of relativistic degrees of freedom is~\cite{Ade:2015xua}
\begin{equation}
N_{\text{eff}} = 3.15 \pm 0.23 \;\; (\text{Planck TT+lowP+BAO}),
\label{neffbound1}
\end{equation}
which is consistent with the standard value $N_{\text{eff}} = 3.046$. 
Here the keywords in parenthesis refer to different constraints imposed to 
obtain the bound, the details of which can be found in Ref.~\cite{Ade:2015xua}. 
The Planck bound is clearly inconsistent with one additional light neutrino. 
Although this latest bound from the Planck experiment can not accommodate one 
additional light sterile neutrino at eV scale within the standard $\Lambda$CDM 
model of cosmology, one can evade 
these tight bounds by considering the presence of some new physics beyond the 
standard model (BSM). 
For example, additional gauge interactions in order to suppress the production 
of sterile neutrinos through flavour oscillations were studied recently by the 
authors of~\cite{Hannestad:2013ana,Dasgupta:2013zpn}. 
Recently, the IceCube experiment at the south pole has excluded the three active 
and one sterile neutrino (the $3+1$ framework where the sterile state is heavier than the active states \cite{GomezCadenas:1995sj,Goswami:1995yq}) parameter space mentioned in 
global fit data~\cite{Kopp:2013vaa} 
at approximately $99\%$ confidence level~\cite{TheIceCube:2016oqi}. 
However, in the presence of non-standard interactions, the $3+1$ neutrino global 
fit data can remain consistent with the IceCube observations~\cite{Liao:2016reh}. 
Therefore, there is still room for existence of an eV scale sterile neutrino 
within some specific BSM frameworks that can provide a consistent interpretation 
of experimental data. 
The interesting cosmological implications of such light sterile neutrinos can be 
found in the recent review article~\cite{Abazajian:2017tcc} and references 
therein.

Apart from finding a consistent $3+1$ neutrino framework compatible with short 
baseline neutrino anomalies as well as cosmology, another challenge in 
particle physics is to 
explain the origin of this light sterile neutrino and its non trivial mixing 
with the active neutrinos. 
Apart from explaining the eV scale mass of sterile neutrino, it is also 
desirable that the particle physics model predicts some of the neutrino 
parameters that 
can undergo further scrutiny at ongoing neutrino oscillation 
experiments~\cite{An:2016luf,MINOS:2016viw,Adamson:2016jku}. 
Typically, a BSM framework for explaining neutrino masses and mixing comes with 
a large number of free parameters lacking predictability. 
However, if the theory has a well motivated underlying symmetry that gives rise 
to a very specific structure of neutrino mass matrix, then number of free 
parameters can be significantly reduced. 
Here we consider such a possibility where an underlying symmetry can restrict 
the mass matrix to have non-zero entries only at certain specific locations. 
Such scenarios are more popularly known as zero texture models, a nice summary 
of which within three neutrino framework can be found in the review article~\cite{Ludl:2014axa}. 
The light neutrino mass matrix in $3+1$ framework is a $4\times 4$ complex 
symmetric matrix, assuming the neutrinos to be Majorana fermions. 
Such a mass matrix can be parametrised by sixteen parameters: four masses, six 
angles and six phases. 
In the presence of zero textures, these parameters get related to 
each other through the zero texture equations resulting in more constrained set 
of parameters or more predictability. 
Recently, the possibilities of such zero textures were explored in the $3+1$ 
framework 
in Ref.~\cite{Ghosh:2012pw,Ghosh:2013nya,Zhang:2013mb,Nath:2015emg,Borah:2016xkc}. 
The authors in these works pointed out the allowed zero texture mass matrices 
containing up to four-zeros in $3+1$ framework from the requirement of satisfying 
recent data of mass-squared differences and mixing angles.
In present paper we briefly summarise all previous works and also extend them to study the 
possibility of having five and six zero texture mass matrices. Since the 
simultaneous existence of zeros 
in active and sterile sectors is phenomenologically disallowed~\cite{Ghosh:2012pw}, 
six is the maximum number of possible zeros in the $4\times4$ 
light neutrino mass matrix. 
Therefore, our present study is going to give a complete picture of all possible 
zero texture mass matrices in $3+1$ framework. 
It should be noted that we stick to a diagonal charged-lepton basis for simplicity 
and hence all our conclusions are valid in this basis only.

After summarising the earlier works on zero texture mass matrices upto 
four-zeros, we show that one possible five-zero texture mass matrix is allowed 
from the present $3+1$ neutrino data while the possibility of six-zero texture 
is ruled out. Apart from finding the predictions for different neutrino 
parameters in this particular five-zero texture mass matrix, we also point out 
one possible symmetry realisation that can naturally generate such a mass 
matrix. This is based on an abelian gauge symmetry where the relative difference 
between second and third generation lepton number $L_{\mu}-L_{\tau}$ is gauged. 
We also discuss other interesting implications of such a scenario related to the 
anomalous magnetic moment of muon. 
We also discuss one interesting discrete symmetry which the five-zero texture 
mass matrix possesses partially and its possible implications.

The paper is organized in the following way. In Section \ref{sec2}, we discuss 
the low energy neutrino mass matrix in the 3+1 framework. 
In Section \ref{sec3} we give a brief summary of the past results on one, two, 
three and
four-zero textures. In Section \ref{sec4} we present our new results on 
five-zero texture. 
Section \ref{sec5} will contain the theoretical model which explain the origin 
of the allowed five-zero texture and Section \ref{sec5b} contains some possible phenomenological implications of our results. Finally we will conclude in Section \ref{sec6}.

\section{Neutrino mass matrix in 3+1 framework}
\label{sec2}

In presence of an extra sterile neutrino having mass in the eV scale, there 
will be two possible mass ordering of the neutrinos: Normal hierarchy (NH) i.e., $m_4 > m_3 > m_2 
> m_1$ and inverted hierarchy (IH) i.e., $m_4 > m_2 > m_1 > m_3$, where
$m_1$, $m_2$, $m_3$ are the masses of the active neutrinos and $m_4$ is the mass 
of the sterile neutrino. They can also have quasidegenerate spectra (QD) if $m_4 > m_3 \sim m_2 
\sim m_1$.
Irrespective of the mass spectrum of the neutrinos, the low energy neutrino mass 
matrix $m_\nu$ in the 3+1
scheme can be expressed as
\begin{eqnarray}
m_{\nu} &=& U m^{\text{diag}}_{\nu} U^T \\
        &=&\begin{pmatrix}
m_{ee} & m_{e\mu} & m_{e\tau} & m_{es}\\
m_{\mu e} & m_{\mu\mu} & m_{\mu\tau} & m_{\mu s} \\
m_{\tau e} & m_{\tau\mu} & m_{\tau\tau} & m_{\tau s} \\
m_{se} & m_{s\mu} & m_{s\tau} & m_{ss}
\end{pmatrix},
\label{mnu}
\end{eqnarray}
where $m_\nu^{\text{diag}} = {\rm diag}(m_1, m_2, m_3, m_4)$ and $U=V.P$ is the 
$4 \times 4$ unitary PMNS matrix which contains six mixing angles i.e., $
\theta_{13}$, $\theta_{12}$, $\theta_{23}$, $\theta_{14}$, $\theta_{24}$,
$\theta_{34}$, three Dirac type CP phases i.e., $\delta_{13}$, $\delta_{14}$, 
$\delta_{24}$ and three Majorana type CP phases i.e., $\alpha$, $\beta$, 
$\gamma$.
$P$ is the diagonal Majorana
phase matrix given by $P = \text{diag} (1, e^{-i\alpha/2}, e^{-i(\beta/2 - 
\delta_{13})}, e^{-i(\gamma/2 - \delta_{14})})$ and we parametrize $V$ as 
\begin{equation}
V = R_{34} \tilde{R}_{24}\tilde{R}_{14}R_{23}\tilde{R}_{13} R_{12},
\end{equation}
%\begin{widetext}
where $R, \tilde{R}$ are the rotation matrices and can be expressed as 
\begin{eqnarray}
R_{34}&=&\begin{pmatrix}
1 & 0 & 0 & 0\\
0 & 1 & 0 & 0 \\
0 & 0 & c_{34} & s_{34} \\
0 & 0 & -s_{34} & c_{34}
\end{pmatrix}, \\
\tilde{R}_{14}&=&\begin{pmatrix}
c_{14} & 0 & 0 & s_{14} e^{-i\delta_{14}}\\
0 & 1 & 0 & 0 \\
0 & 0 & 1 & 0 \\
-s_{14} e^{i\delta_{14}} & 0 & 0 & c_{14} 
\end{pmatrix},
\end{eqnarray}
and so on, with $c_{ij} = \cos{\theta_{ij}}, \; s_{ij} = \sin{\theta_{ij}}$ and 
$\delta_{ij}$ are the Dirac CP phases. In this parametrization, the 
six CP phases vary from $-\pi$ to $\pi$. 

% The latest information of the oscillation parameters for the standard three 
%flavor 
% i.e., $\theta_{13}$, $\theta_{12}$, $\theta_{23}$, $\Delta m^2_{21}$, $|\Delta 
%m^2_{\rm atm}|$ ($\Delta m^2_{31}$ for NH and $\Delta m^2_{32}$ for IH)  
% and $\delta_{13}$ come from the global analysis of the data
% coming from various accelerator, reactor and atmospheric neutrino oscillation 
%experiments~\cite{Forero:2014bxa,Esteban:2016qun,Capozzi:2013csa}.
% The information for the sterile mixing angles come from....
% {\bf Suprabh: add a discussion on the sterile mixing parameters}

\begin{table*}
\begin{small}
\begin{tiny}
\begin{center}
\begin{small}
\begin{tabular}{|c|c|c|c|}
\hline $ A_1$& $A_2$ &  &   \\
\hline $\left(
\begin{array}{cccc}
0 & 0 & \times &\times \\  0 & \times & \times & \times \\ \times & \times & 
\times & \times \\\times & \times & \times & \times
\end{array}
\right)$ & $\left(
\begin{array}{cccc}
0 & \times & 0 &\times \\  \times & \times & \times & \times \\ 0 & \times & 
\times & \times \\\times & \times & \times & \times
\end{array}
\right)$  & & \\
\hline $ B_1$ & $B_2$ & $B_3$ & $B_4$  \\
\hline
 $\left(
\begin{array}{cccc}
\times & \times & 0 &\times \\  \times &0 & \times & \times \\ 0 & \times & 
\times & \times \\\times & \times & \times & \times
\end{array}
\right)$  & $\left(
\begin{array}{cccc}
\times& 0 & \times &\times \\  0 &\times & \times & \times \\ \times & \times & 
0& \times \\\times & \times & \times & \times
\end{array}
\right)$& $\left(
\begin{array}{cccc}
\times & 0 & \times &\times \\  0 & 0 & \times & \times \\ \times & \times & 
\times & \times \\\times & \times & \times & \times
\end{array}
\right)$  & $\left(
\begin{array}{cccc}
\times & \times& 0 &\times \\  \times & \times & \times & \times \\0 & \times & 
0 & \times \\\times & \times & \times & \times
\end{array}
\right)$ \\
\hline $C$& & &  \\
\hline
 $\left(
\begin{array}{cccc}
\times & \times & \times &\times \\ \times & 0 & \times & \times \\ \times & 
\times & 0 & \times \\\times & \times & \times & \times
\end{array}
\right)$ & & & \\
\hline $D_1$& $D_2$ & & \\
\hline
$\left(
\begin{array}{cccc}
\times &\times & \times &\times \\  \times & 0 & 0 & \times \\ \times & 0 & 
\times & \times \\\times & \times & \times & \times
\end{array}
\right)$ & $\left(
\begin{array}{cccc}
\times & \times & \times &\times \\  \times & \times &0 & \times \\ \times & 0 
&0 & \times \\\times & \times & \times & \times
\end{array}
\right)$& & \\
\hline $E_1$ & $E_2$ & $E_3$ & \\
\hline
$\left(
\begin{array}{cccc}
0 & \times & \times &\times \\ \times & 0 & \times & \times \\ \times & \times & 
\times & \times \\\times & \times & \times & \times
\end{array}
\right)$& $\left(
\begin{array}{cccc}
0 & \times & \times &\times \\  \times & \times & \times & \times \\ \times & 
\times & 0 & \times \\\times & \times & \times & \times
\end{array}
\right)$&  $\left(
\begin{array}{cccc}
0 &\times & \times &\times \\ \times & \times & 0& \times \\ \times & 0 & \times 
& \times \\\times & \times & \times & \times
\end{array}
\right)$& \\
\hline $F_1$& $F_2$ & $F_3$ & \\
\hline
$\left(
\begin{array}{cccc}
\times & 0 & 0 &\times \\  0 & \times & \times & \times \\0 & \times & \times & 
\times \\\times & \times & \times & \times
\end{array}
\right)$&  $\left(
\begin{array}{cccc}
\times& 0 & \times &\times \\  0 & \times & 0 & \times \\ \times & 0 & \times & 
\times \\\times & \times & \times & \times
\end{array}
\right)$ & $\left(
\begin{array}{cccc}
\times &\times & 0 &\times \\ \times & \times &0 & \times \\ 0 & 0 & \times & 
\times \\\times & \times & \times & \times
\end{array}
\right)$&\\
\hline
\end{tabular}
\end{small}
\end{center}
\end{tiny}
\caption{Possible two-zero textures in $m_{\nu}$ in the 3+1 scenario.}
\label{tab1}
\end{small}
\end{table*}

\section{Previous results of zero textures in $m_{\nu}$ in 3+1 scheme}
\label{sec3}

In this section we will discuss briefly the previous results of one, two, three 
and four-zero textures.
%\subsection{one-zero texture}
One-zero texture in the neutrino mass matrix is given by the condition
\begin{eqnarray}
 m_{\alpha \beta} = 0
\end{eqnarray}
where $\alpha$, $\beta$ are the flavour indices. Thus there exist ten possible 
one-zero mass matrices. One-zero mass matrices in 3+1 scheme have been discussed in 
Refs.~\cite{Ghosh:2013nya,Nath:2015emg}.
%Note that understanding of one-zero textures is extremely important because 
%with the knowledge of one-zero texture one can understand the properties of two, 
%three and four-zero textures. As the
%analytical expression of the elements of $m_\nu$ are quite complicated in 3+1 
%scheme it is almost impossible to understand the properties of two, three and 
%four-zero texture analytically.
%Thus once the parameter dependence of the one zero texture is understood it is 
%easy to infer the parameter correlations for different two, three and four zero 
%textures.
The main results of these works are:
\begin{itemize}
 \item All the one-zero textures except $m_{es}=0$ and $m_{ss}=0$ are allowed.
 \item The texture $m_{ee}=0$ is allowed in both NH and IH. This is in sharp 
contrast to the standard three flavour case where $m_{ee} = 0$ is not allowed in 
IH. 
 Note that the texture zero condition of $m_{ee} = 0$ in 3+1 scenario depends on 
the value of $\theta_{14}$. 
 Specifically in this case $\theta_{14}$ is a rising function of the lowest 
neutrino mass. 
% It has been found that if very small values of $\theta_{14}$ are
% not allowed (i.e., $\theta_{14}$ close to zero), then it is not possible to 
%have $m_{ee}=0$ for NH when the lowest neutrino mass is zero. 
% But on the other hand for IH one does not need very small values of 
%$\theta_{14}$ to get $m_{ee} = 0$ even when $m_3$ is zero. 
 The condition $m_{ee} = 0$ also strong constrains
 Majorana phase $\gamma$ to be around $\pm \pi$. Thus any two, three and 
four-zero texture that 
 involve $m_{ee} = 0$ will predict $\theta_{14}$ as a rising function of lowest 
mass and $\gamma$ around $\pm \pi$.
 
 \item The textures $m_{e\mu}=0$, $m_{e\tau}=0$, $m_{\mu \mu}=0$, 
$m_{\mu\tau}=0$ and $m_{\tau \tau}=0$ are allowed in both NH and IH.
 \item The texture $m_{\mu s}=0$ is only allowed if the neutrino masses are 
quasi-degenerate.
\end{itemize}
Before moving forward let us discuss a bit more about the elements in the fourth 
row/column of $m_\nu$. For the elements $m_{es}$, $m_{\mu s}$, $m_{\tau s}$ and 
$m_{ss}$,
the leading order term looks like $\sim m_4 s_{14}$, $\sim m_4 s_{24}$, $\sim m_4 
s_{34}$ and  $\sim m_4$ respectively. Now as $m_4$ is quite large ($\Delta 
m^2_{14} \sim 1$ eV$^2$), the 
coefficient of $m_4$ needs to be small to obtain $m_{\alpha s} = 0 $ (with 
$\alpha = e$, $\mu$, $\tau$ and $s$). Thus we see that $m_{ss}$ 
can never be zero. In the analysis of
Refs.~\cite{Ghosh:2013nya,Nath:2015emg}, the mixing angles $\theta_{14}$ and 
$\theta_{24}$ are bounded from below but $\theta_{34}$ can be as small as 
zero. This is the reason why they have
concluded that $m_{es}=0$ is not allowed and $m_{\mu s }$ vanishes only in the 
quasi-degenerate regime. But if one assumes the values of $\theta_{14}$ and 
$\theta_{24}$ close to zero are allowed,
then the conclusions about $m_{es}$ and $m_{\mu s}$ may change.

%\subsection{Two-zero texture}
Two-zero textures in neutrino mass matrix are obtained when two of the matrix 
elements are zero simultaneously. In 3+1 scheme the two-zero textures are 
discussed in Ref.~\cite{Ghosh:2012pw}.
The number of possible two-zero textures in $m_{\nu}$ is 45. Among the 45 cases, 
there are 30 cases in which the zero texture includes $m_{\alpha s} = 0$ (where 
$\alpha = e$, $\mu$, $\tau$ and $s$).
It was shown that any texture zero which includes an element 
corresponding to the fourth row or fourth column of the mass matrix is not allowed \footnote{In this analysis, the mixing angles $\theta_{14}$ and $\theta_{24}$ are considered to be bounded from below.
However, numerical analysis reveals that even when $\theta_{14}$ and $\theta_{24}$ are close to zero, the two-zero textures involving the elements corresponding to the fourth row or fourth column are not allowed.}. 
Thus we are left with the 15 two-zero textures
listed in Table \ref{tab1}.
Here it is interesting to note that these 15 textures coincide with the 15 
possible two-zero textures in the standard three generation in the absence of 
the sterile neutrino.
There are numerous studies in the literature which discuss the viability of 
these 15 two-zero textures in 3 
generation~\cite{Xing:2004ik,onezero,onezero1,onezero2,onezero3,alltex,texturesym9,twozero,
twozero1,twozero2,twozero3,twozero4,twozero4_1,
twozero5,twozero6,twozero7,twozero8,twozero9,twozero10,twozero11,Gautam:2016qyw}.
All these analyses show that among these 15 two-zero textures in three generations, 
only seven textures are allowed. These allowed textures belong to $A$, $B$ and 
$C$ class as given in Table \ref{tab1}.
But the conclusions in these studies change when one includes a light sterile neutrino in addition 
to the three active neutrinos. The analysis of Ref.~\cite{Ghosh:2012pw} shows that in 
the 3+1 scheme
only the textures in class $A$ are allowed in NH whereas textures belonging to all 
the other classes (i.e., $B$, $C$, $D$, $E$ and $F$) are allowed in 
both NH and IH. This is the most remarkable finding of this work. The textures 
which were disallowed in the three-generation case become allowed from the 
contribution of 
additional terms from the sterile sector.

The analysis of zero textures when three elements of the neutrino mass matrix 
are simultaneously zero can be found in Ref.~\cite{Zhang:2013mb}.
Note than in the standard three-generation case the maximum allowed numbers of zero 
textures are two while three-zero textures are phenomenologically disallowed. Thus the possibility 
of having more than two zeros in $m_{\nu}$ is
a special feature of the 3+1 scheme.
In the 3+1 scenario, there can be 120 possible three-zero
textures. But among them 100 three-zero textures contain an element belonging 
to the fourth row/fourth column of the mass matrix and hence they are not 
allowed. 
%\st{Among the remaining 20 textures
%it was found that 19 are in accordance with the neutrino oscillation data.} 
The 20 textures are classified in six sets as follows:
\begin{widetext}
\begin{eqnarray}
   A &:&
  \begin{pmatrix} 
  0 & \times & \times & \times \\ 
  \times  & 0 & \times & \times \\  
  \times & \times & 0 & \times \\ 
  \times  & \times & \times & \times  
  \end{pmatrix}; \\ \nonumber
   B_1 : \begin{pmatrix} 
   0 & \times & 0 & \times \\ 
   \times  & 0 & \times & \times \\ 
   0 & \times & \times & \times \\  
   \times & \times & \times & \times  
   \end{pmatrix},
   B_2 &:&
   \begin{pmatrix}
 0 & \times & \times & \times \\ 
 \times & 0 & 0 & \times \\ 
 \times & 0 & \times & \times \\  
 \times & \times & \times & \times 
 \end{pmatrix},
   B_3 : \begin{pmatrix}
 0 & 0 & \times & \times \\ 
 0  & \times & \times & \times \\ 
 \times & \times & 0 & \times \\  
 \times & \times & \times & \times  
 \end{pmatrix}; \\ \nonumber
   B_4 :\begin{pmatrix} 
0 & \times & \times & \times \\  
\times & \times & 0 & \times \\ 
\times & 0 & 0 & \times \\ 
\times & \times & \times & \times  
\end{pmatrix},
   B_5 &:& \begin{pmatrix} 
\times & 0 & \times & \times \\ 
0  & 0 & \times & \times \\ 
\times &\times& 0 & \times \\  
\times&\times&\times& \times  
\end{pmatrix},
  B_6 : \begin{pmatrix} 
\times & \times & 0 & \times \\  
\times & 0 & \times & \times \\ 
0 &\times& 0 & \times \\ 
\times &\times&\times& \times  
\end{pmatrix}; \\ \nonumber
  C_1 : \begin{pmatrix} 
0 & 0 & \times & \times \\ 
0  & 0 & \times & \times \\ 
\times &\times& \times & \times \\ 
\times &\times&\times& \times  
\end{pmatrix},
   C_2 &:& \begin{pmatrix} 
0 & \times & 0 & \times \\ 
\times & \times & \times & \times \\ 
0 &\times& 0 & \times \\  
\times &\times&\times& \times 
\end{pmatrix},
   C_3 : \begin{pmatrix}
 \times & \times & \times & \times \\  
 \times & 0 & 0 & \times \\ 
 \times &0& 0 & \times \\  
 \times&\times&\times& \times  
 \end{pmatrix}; \\ \nonumber
  D_1 : \begin{pmatrix} 
 0 & 0 & \times & \times \\ 
 0  & \times & 0 & \times \\ 
 \times &0& \times & \times \\  
 \times&\times&\times& \times  
 \end{pmatrix},
   D_2 &:& \begin{pmatrix} 
0 & \times & 0 & \times \\ 
\times & \times & 0 & \times \\ 
0 &0& \times & \times \\   
\times&\times&\times& \times 
\end{pmatrix},
   D_3 : \begin{pmatrix} 
\times & 0 & 0 & \times \\ 
0  & 0 & \times & \times \\ 
0 &\times& \times & \times \\  
\times&\times&\times& \times  
\end{pmatrix}; \\ \nonumber
 D_4 : \begin{pmatrix} 
 \times & \times & 0 & \times \\ 
 \times  & 0 & 0 & \times \\ 
 0 &0& \times & \times \\  
 \times&\times&\times& \times  
 \end{pmatrix},
   D_5 &:& \begin{pmatrix} 
\times & 0 & 0 & \times \\  
0 & \times & \times & \times \\ 
0 &\times& 0 & \times \\  
\times&\times&\times& \times  
\end{pmatrix},
   D_6 : \begin{pmatrix}
 \times & 0 & \times & \times \\ 
 0 & \times & 0 & \times \\ 
 \times &0& 0 & \times \\ 
 \times &\times&\times& \times  
 \end{pmatrix}; \\ \nonumber
  E_1 : \begin{pmatrix} 
0 & 0 & 0 & \times \\  
0 & \times & \times & \times \\ 
0 &\times& \times & \times \\ 
\times &\times&\times& \times  
\end{pmatrix},
   E_2 &:& \begin{pmatrix}
 \times & 0 & \times & \times \\ 
 0 & 0 & 0 & \times \\ 
 \times &0& \times & \times \\  
 \times &\times&\times& \times 
 \end{pmatrix},
   E_3 : \begin{pmatrix}
 \times & \times & 0 & \times \\  
 \times & \times & 0 & \times \\ 
 0 &0& 0 & \times \\  
 \times&\times&\times& \times  
 \end{pmatrix}; \\ \nonumber
   F &:& \begin{pmatrix}
 \times & 0 & 0 & \times \\ 
 0  & \times & 0 & \times \\ 
 0 & 0 & \times & \times \\  
 \times&\times&\times& \times  
 \end{pmatrix}.
\end{eqnarray}
\end{widetext}
Numerical analysis shows that 
%\st{apart from the texture $C_3$, all the 19 textures 
%are viable in this case.} 
the textures $A$, $B_2$, $B_4$, $B_5$, $B_6$, $D_3$, 
$D_4$, $D_5$, $D_6$, $E_2$ and $E_3$
are allowed in both NH and IH whereas the remaining seven textures i.e., $B_1$, 
$B_3$, $C_1$, $C_2$, $D_1$, $D_2$ and $E_1$ prefer NH over IH. 
The texture $C_3$ is excluded almost completely, being allowed in a very small 
part of the parameter space for IH.

An analysis of four-zero textures, when four elements of the low energy 
Majorana neutrino mass matrix can vanish simultaneously is carried out in Ref.~\cite{Borah:2016xkc}.
In this case the number of possible zero textures are 210. But again out of 
these 210 textures, 195 are readily ruled out because for these textures we have 
$m_{\alpha s} = 0$.
The remaining 15 textures are then classified in either group $A$ where the 
texture zeros contain $m_{ee} = 0$ or group $B$ in which $m_{ee} \neq 0$.
\begin{widetext}
\begin{eqnarray}
A_1 &:&\begin{pmatrix}
0& 0 &\times & \times\\
 0 & 0 &\times & \times \\
\times& \times & 0 &\times \\
\times & \times & \times & \times 
\end{pmatrix} , 
  A_2 :\begin{pmatrix}
0& \times &0 & \times\\
\times& 0 &\times & \times \\
0 & \times & 0 &\times \\
\times & \times & \times & \times 
\end{pmatrix},
  A_3 :\begin{pmatrix}
0& \times &\times & \times\\
\times& 0 & 0 & \times \\
\times& 0 & 0 &\times \\
\times & \times & \times & \times 
\end{pmatrix};  \\ \nonumber
A_4 &:&\begin{pmatrix}
0& 0 &0 & \times\\
0 & 0 &\times & \times \\
 0 & \times & \times &\times \\
\times & \times & \times & \times 
\end{pmatrix} , 
  A_5 :\begin{pmatrix}
0& 0 &\times & \times\\
0 & 0 & 0 & \times \\
\times& 0 & \times &\times \\
\times & \times & \times & \times 
\end{pmatrix},
  A_6 :\begin{pmatrix}
0& \times &0 & \times\\
\times & 0 & 0 & \times \\
0 & 0 & \times &\times \\
\times & \times & \times & \times 
\end{pmatrix}, \\ \nonumber
A_7 &:&\begin{pmatrix}
0& 0 &0 & \times\\
0 & \times &\times & \times \\
 0 & \times & 0 &\times \\
\times & \times & \times & \times 
\end{pmatrix} , 
  A_8 :\begin{pmatrix}
0& 0 &\times & \times\\
0 & \times & 0 & \times \\
\times& 0 & 0 &\times \\
\times & \times & \times & \times 
\end{pmatrix},
  A_9 :\begin{pmatrix}
0& \times &0 & \times\\
\times & \times & 0 & \times \\
0 & 0 & 0 &\times \\
\times & \times & \times & \times 
\end{pmatrix}, \\ \nonumber
 A_{10}&:&\begin{pmatrix}
0 & 0 & 0 & \times \\
0 & \times &0 & \times \\
 0 & 0 & \times &\times \\
\times & \times & \times & \times
\end{pmatrix},
 B_1 :\begin{pmatrix}
\times& 0 &\times & \times\\
0 & 0 & 0 & \times \\
\times& 0 & 0 &\times \\
\times & \times & \times & \times 
\end{pmatrix},
  B_2 :\begin{pmatrix}
\times& \times &0 & \times\\
\times & 0 & 0 & \times \\
0 & 0 & 0 &\times \\
\times & \times & \times & \times 
\end{pmatrix}, \\ \nonumber
 B_3 &:&\begin{pmatrix}
\times & 0 & 0 & \times \\
0 & 0 &\times & \times \\
 0 & \times & 0 &\times \\
\times & \times & \times & \times  
\end{pmatrix} ,  
  B_4 :\begin{pmatrix}
\times & 0 & 0 & \times \\
0 & 0 &0 & \times \\
 0 & 0 & \times &\times \\
\times & \times & \times & \times 
\end{pmatrix},
  B_5 :\begin{pmatrix}
\times & 0 & 0 & \times \\
0 & \times &0 & \times \\
 0 & 0 & 0 &\times \\
\times & \times & \times & \times 
\end{pmatrix}.
\end{eqnarray}
\end{widetext}
Numerical analysis reveals that the textures $A_3$, $B_1$ and $B_2$ are the 
disallowed in NH whereas $A_9$ and $A_{10}$
are the disallowed in IH. 

Note that results of the one, two, three and four-zero neutrino mass matrix textures are largely consistent 
with each other. In Ref.~\cite{Ghosh:2012pw}, it was shown that the two-zero textures 
$A_1$ and $A_2$ are disallowed
in IH. In class $A_1$ the elements $m_{ee}$ and $m_{e \mu}$ vanish whereas in 
class $A_2$ we have $m_{ee} = m_{e\tau}=0$. Thus this work 
predicts any simultaneous zero texture
involving $m_{e \mu}$ or $m_{e\tau}$ with $m_{ee}$ will be disallowed in IH. In 
the analysis of three-zero textures, the classes where we obtain  $m_{ee} = m_{e 
\mu} = 0$ or $m_{ee} = m_{e\tau} = 0$
are $B_1$, $B_3$, $C_1$, $C_2$, $D_1$, $D_2$ and $E_1$. According to the 
analysis of Ref.~\cite{Zhang:2013mb} though these cases are preferred in NH over 
IH, they are not completely ruled out in IH.
This difference is mainly due to the choice of different ranges
 of the sterile mixing parameters, in particular $\theta_{14}$,
  $\theta_{24}$ and $\Delta m^2_{41}$. In the four-zero case
   the textures where
we have the condition $m_{ee} = m_{e \mu} = 0$ or $m_{ee} = m_{e\tau} = 0$ are 
$A_1$, $A_2$, $A_4$, $A_5$, $A_6$, $A_7$, $A_8$, $A_9$ and $A_{10}$. 
According to the analysis of Ref.~\cite{Borah:2016xkc}, all other textures except $A_9$ and 
$A_{10}$ are allowed in IH. This difference occurs because in Ref.~\cite{Borah:2016xkc},
 there are no lower
bounds on $\theta_{14}$ and $\theta_{24}$ and these can be as low as zero.
% \st{Another difference is the three-zero texture $C_3$
%   and the four
% zero texture $A_3$ both of which contains the condition
%  $m_{\mu\mu} = m_{\mu \tau} = m_{\tau\tau} = 0$. According to} Ref.~\cite{Zhang:2013mb},
%  \st{$C_3$ is disallowed in both NH and IH but according to }
%  Ref.~\cite{Borah:2016xkc} \st{$A_3$ is disallowed in NH but allowed in IH. If we 
% look at the lower right panel of Fig. 5 of} Ref.~\cite{Zhang:2013mb}, \st{then we see 
% that $C_3$ is marginally allowed in IH.
% It is important to note that the input ranges of neutrino parameters in both the 
% studies are quite different. In} Ref.~\cite{Zhang:2013mb}, \st{the three complex 
% equations corresponding to three-zero conditions 
% are solved in terms of three mass ratios $m_2/m_1$, $m_3/m_1$ and $m_4/m_1$. 
% Whereas in} Ref.~\cite{Borah:2016xkc}, \st{eight real equations corresponding to 
% the four-zero texture are solved in terms of eight 
% knowns $m_1$ ($m_3$), $\theta_{34}$ and six phases for NH (IH) while using the 
% other eight parameters as input.}
Thus we understand that the viability of zero textures in 3+1 case is extremely 
sensitive to the choice of the ranges of active-sterile mixing parameters. 
%\st{and also depends on the method of scanning the parameter 
%space}.

\section{Results of five-zero textures}
\label{sec4}

\begin{figure*}
\begin{center}
\includegraphics[width=0.4\textwidth]{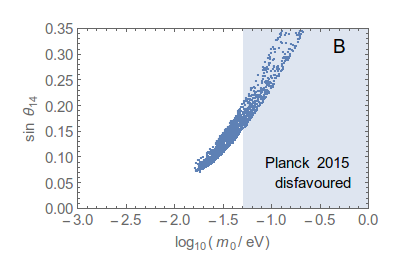}
\includegraphics[width=0.4\textwidth]{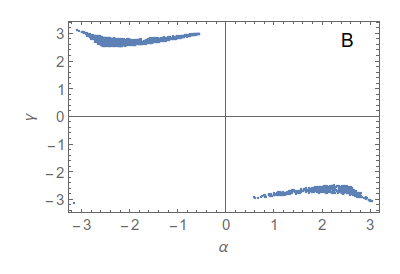} \\
\includegraphics[width=0.4\textwidth]{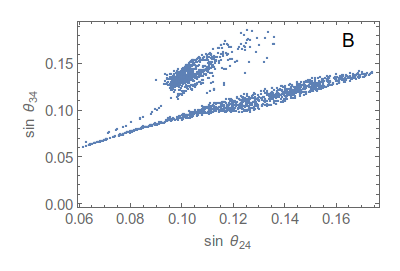}
\includegraphics[width=0.4\textwidth]{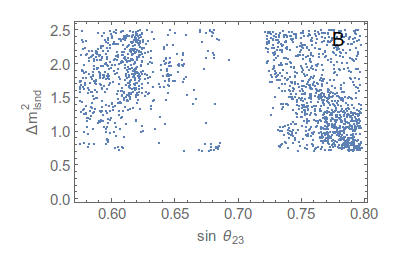}
\end{center}
\begin{center}
\caption{Correlation plots of class $B$ in NH. $m_0$ is the lowest neutrino mass 
which is $m_1$ in NH.}
\label{fig1}
\end{center}
\end{figure*}

In this section we present our results for the five-zero textures. The five-zero 
texture condition can be expressed mathematically as
\begin{eqnarray}
 a_i m_1 + b_i m_2 + c_i m_3 + d_i m_4 = 0, \quad (i \in \{1-5\})
 \label{eq1}
\end{eqnarray}
where $a_i$, $b_i$, $c_i$ and $d_i$ are functions of the mixing angles and 
phases. Note that Eq. \ref{eq1} is a set of five complex equations. To obtain 
the real set of equations we
put the real part and the imaginary part individually to zero:
\begin{eqnarray}
  a_i^\prime m_1 + b_i^\prime m_2 + c_i^\prime m_3 + d_i^\prime m_4 = 0, \quad 
(i \in \{1-10\}).
  \label{eq2}
 \end{eqnarray}
 %%%%%%%%%%%%%%%%%%%%%%%%%%%%%%%%%%%%%%%%%%%%%%%%%%%%%%%%%%
\begin{table*}
\begin{center}
\begin{tabular}{|c|c|c|c|}
\hline Parameters & Allowed ranges & Parameters & Allowed ranges\\
\hline
$ \theta_{12}$ & 3 $\sigma$ & $\delta_{13}$ & -180$^\circ$ to 180$^\circ$ \\
\hline
 $\theta_{13}$ & 3 $\sigma$  & $\delta_{14}$ & -86$^\circ$ to 86$^\circ$ \\
\hline
 $\theta_{23}$ & $\neq$ 45$^\circ$ & $\delta_{24}$ & -180$^\circ$ and 
180$^\circ$\\
\hline 
$\theta_{14}$ & $>$ 3$^\circ$ & $\alpha$ & -180$^\circ$ to -30$^\circ$\\
 & & & 30$^\circ$ to 180$^\circ$\\
\hline
$\theta_{24}$ & $>$ 3$^\circ$ & $\beta$ & -35.5$^\circ$ to 35.5$^\circ$\\
\hline 
$\theta_{34}$ &  3.4$^\circ$ - 11$^\circ$ & $\gamma$ & -180$^\circ$ to 
-137$^\circ$ \\
& & & 137$^\circ$ to 180$^\circ$\\
\hline
$m_1$ & 0.018 - 0.22 eV & $\Delta{m^2_{LSND}} $ & 0.7 - 2.5 eV$^2$\\
\hline
\end{tabular}	
\end{center}
\caption{Allowed ranges of the neutrino oscillation parameters for texture $B$.}
\label{tab2}
\end{table*} 
 %%%%%%%%%%%%%%%%%%%%%%%%%%%%%%%%%%%%%%%%%%%%%%%%%%%%%%%%%%
Now Eq.~\ref{eq2} is a set of ten real equations relating sixteen independent 
parameters i.e. six mixing angles, six phases and four masses. 
To solve this set of equations we 
supply the input values of three active neutrino mixing angles (i.e., 
$\theta_{12}$, $\theta_{13}$ and $\theta_{23}$), and
three mass square differences (i.e., $\Delta m^2_{21}$, $|\Delta m^2_{31}|$ and 
$\Delta m^2_{\rm LSND}$). The ten equations are solved for the remaining ten 
parameters using Mathematica (which internally implements the multi-dimensional 
Newton-Raphson algorithm).
For NH, we have expressed $m_2$, $m_3$ and $m_4$ as $\sqrt{m^2_1+\Delta 
m_{21}^2}$, $\sqrt{m_1^2+\Delta m_{31}^2}$ and $\sqrt{m_1^2+\Delta m_{41}^2}$ 
respectively, 
keeping the lowest mass $m_1$ free, 
whereas for IH we have expressed $m_1$, $m_2$ and $m_4$ as $\sqrt{m_3^2-\Delta 
m_{32}^2-\Delta m_{21}^2}$, $\sqrt{m_3^2-\Delta m_{32}^2}$ and 
$\sqrt{m_3^2+\Delta m_{43}^2}$ 
keeping the lowest mass $m_3$ free.
We have varied our input parameters for the three-generation parameters
in the $3 \sigma$ allowed range as given the global analysis of the world 
neutrino data~\cite{Forero:2014bxa,Esteban:2016qun,Capozzi:2013csa} 
and varied $\Delta m^2_{\rm LSND}$ from 0.7 eV$^2$ to 2.5 eV$^2$. 
If the output of $\theta_{14}$, $\theta_{24}$ and $\theta_{34}$ falls between 
$0^\circ$ to $20^\circ$, $0^\circ$ to $11.5^\circ$ and $0^\circ$ to $30^\circ$ 
respectively~\cite{Kopp:2013vaa,An:2014bik,Adamson:2011ku} with the condition $m_1 (m_3) > 
0$, then we say this texture is allowed in NH (IH)\footnote{Note that according to the global analysis of the short-baseline data \cite{Kopp:2013vaa} we have $6^\circ < \theta_{14} < 20^\circ$ and
$3^\circ < \theta_{24} < 11.5^\circ$ at $3 \sigma$. However the Refs. \cite{An:2014bik,Adamson:2011ku}, give only an upper limit on $\theta_{14}$ and $\theta_{24}$
as they analyse
stand-alone data. Thus for a conservative approach, in our analysis we have taken the upper limits of $\theta_{14}$ and $\theta_{24}$ from the global analysis and allowed them to have lower limits as zero.}. 
In 3+1 scenario, the number 
of possible five-zero textures are 
$^{10}C_5 = 252  $. But among them, 246 appears with the with one of the 
elements belonging to the fourth row/column and thus they are not allowed.
The remaining six possible five-zero textures are:
\begin{widetext}
\begin{eqnarray}
A &:&\begin{pmatrix}
0 & 0 & 0 & \times\\
 0 & 0 & 0 & \times \\
0 & 0 & \times &\times \\
\times & \times & \times & \times 
\end{pmatrix} , 
  B :\begin{pmatrix}
0 & 0 &0 & \times\\
0 & 0 &\times & \times \\
0 & \times & 0 &\times \\
\times & \times & \times & \times 
\end{pmatrix},
  C :\begin{pmatrix}
0& 0 & 0 & \times\\
0 & \times & 0 & \times \\
0 & 0 & 0 &\times \\
\times & \times & \times & \times 
\end{pmatrix};  \\ \nonumber
D &:&\begin{pmatrix}
0 & 0 & \times & \times\\
0 & 0 & 0 & \times \\
\times & 0 & 0 &\times \\
\times & \times & \times & \times 
\end{pmatrix} , 
  E :\begin{pmatrix}
0& \times & 0 & \times\\
\times & 0 & 0 & \times \\
0 & 0 & 0 &\times \\
\times & \times & \times & \times 
\end{pmatrix},
  F :\begin{pmatrix}
 \times & 0 &0 & \times\\
0 & 0 & 0 & \times \\
0 & 0 & 0 &\times \\
\times & \times & \times & \times 
\end{pmatrix}. \nonumber
\end{eqnarray}
\end{widetext}
Among these six possible structures, our analysis shows that only the texture $B$ 
is allowed in NH, and all the textures are ruled out in IH. 
In Fig. \ref{fig1}, we have given the correlation plots for the allowed texture 
$B$ in NH. In the texture $B$, we have the condition $m_{ee} = 0$. As mentioned 
in the previous section,
the property that any zero texture mass matrix having the condition $m_{ee} = 
0$, will have $\theta_{14}$ as a rising function of the lowest mass and 
the Majorana phase will be constrained around $\pm \pi$, as clearly seen in 
Fig.~\ref{fig1} (top right panel). This property can be simply understood by looking at 
the expression of $m_{ee}$.
\begin{eqnarray}
\label{m}
 m_{ee} &=& c_{12}^2 c_{13}^2 c_{14}^2 m_1+e^{- i \alpha } c_{13}^2 c_{14}^2 m_2 
s_{12}^2 \\ \nonumber
 &+& e^{- i \beta } c_{14}^2 m_3 s_{13}^2+e^{- i \gamma } m_4 s_{14}^2.
\end{eqnarray}
Note that in the above equation the sterile term is given by $ e^{- i \gamma 
} m_4 s_{14}^2$. Now the condition of $m_{ee} = 0$ is simply obtained by the cancellation 
of active and sterile terms.
Therefore it is easy to understand that when $m_1$ is small (large) we need 
smaller (larger) values of $\theta_{14}$ to achieve cancellation. At the same 
time, for the cancellation of 
sterile and active terms, the coefficient of the sterile term must 
acquire a negative sign which is only possible if the phase $\gamma$ is around 
$\pm \pi$.
From Eq.~\ref{m} we also infer that $\theta_{14} \to 0$ leads to the 
standard three-flavor neutrino mixing scenario.
In that case the lowest mass $m_1$ cannot be zero in order
to produce zero texture at $m_{ee}$. The vanishing of the 
active-sterile mixing angles results in the reduction of expressions
 of $m_{ee}$, $m_{\mu\mu}$ and $m_{\tau\tau}$ from four-neutrino
 mixing to three-neutrino mixing scenarios. Earlier studies~\cite{Ghosh:2013nya, Nath:2015emg} 
 have shown that
  the active-sterile mixing angles $\theta_{14}$, $\theta_{24}$ and
  $\theta_{34}$ and the lowest mass $m_1$ cannot simultaneously
 vanish in order to produce zeros
 textures $m_{ee}$, $m_{\mu\mu}$ and 
$m_{\tau\tau}$. We summarize the allowed ranges of neutrino
 oscillation parameters in Table~\ref{tab2} for texture $B$. As
seen from Fig.~\ref{fig1}, very small $\theta_{14}$ and
 $m_1$ are disallowed for the current texture under consideration (top left panel). 
The same figure also shows the constraint on the 
lightest neutrino mass from the upper bound on the sum of absolute neutrino 
masses $\sum_i \lvert m_i \rvert < 0.17$ eV given by the latest data from the 
Planck mission~\cite{Ade:2015xua}. The bottom left panel shows that 
$\theta_{34}<11^\circ$. This can be attributed to the fact that 
a very large value
 of $\theta_{34}$ negates the existence of zero texture at $m_{\tau\tau}$. A 
rigorous analysis of zero textures
at each element in
neutrino mass matrix
 and the interdependency of neutrino parameters is done in Refs.~\cite{Ghosh:2013nya,
 Nath:2015emg}, and their results apply here. From the bottom right panel 
 of Fig.~\ref{fig1}, we also see that 
the maximal value of $\theta_{23}$ is disallowed in this texture. Note that the 
features discussed above are of great importance to
 probe this texture in the future generation oscillation experiments. 
 For example, if the future experiments measure $\theta_{14}$ or $\theta_{24} < 
3^\circ$ or $\theta_{34}$ not in the region $3.4^\circ$ to $11^\circ$ or 
$\theta_{45} = 45^\circ$, then 
 this texture can be readily ruled out
 and hence the possibility of having 5 zeros in $m_{\nu}$.
 
From the earlier results it is obvious that five-zero textures are the 
maximum which is allowed in the 3+1 scheme. 
A texture containing more than five-zeros in the low energy neutrino mass matrix 
is not allowed in the 3+1 scenario. 

\section{Flavor symmetry origin of five-zero texture}
\label{sec5}

\begin{table}
\begin{center}
\begin{tabular}{|c|c|c|c|}
\hline
Fermion Fields & $SU(3)_c \times SU(2)_L \times U(1)_Y$ & 
$U(1)_{L_{\mu}-L_{\tau}}$ & $U(1)_S$ \\
\hline
$ L_e$ & $ (1, 2, -1)$ & $0$ & $0$\\
$ L_\mu$ & $ (1, 2, -1)$ & $1$ & $0$\\
$ L_\tau$ & $ (1, 2, -1)$ & $-1$ & $0$\\
$ \nu_s $ & $ (1, 1, 0)$ & $0$ & $n$\\
$ N_{\mu} $ & $ (1, 1, 0)$ & $1$ & $0$\\
$ N_{\tau} $ & $ (1, 1, 0)$ & $-1$ & $0$\\
\hline
Scalar Fields & $SU(3)_c \times SU(2)_L \times U(1)_Y$ & 
$U(1)_{L_{\mu}-L_{\tau}}$ & $U(1)_S$\\
\hline
$ H_1 $ & $(1,2,1)$ & $0$ & $0$\\
$ H_2 $ & $(1,2,1)$ & $0$ & $n$\\
$ \chi_1 $ & $(1,1,0)$ & $-1$ & $-n$\\
$ \chi_2 $ & $(1,1,0)$ & $1$ & $-n$\\
\hline
\end{tabular}
\end{center}
\caption{Fields responsible for $4\times4$ light neutrino mass matrix with 
five-zero texture}
\label{tab3}
\end{table}
Since the zero textures appear only in the active $3\times3$ block of  the $4\times4$ 
mass matrix, they can be explained by 
different flavor symmetry frameworks. Some possible models are discussed in 
Refs.~\cite{texturesym,texturesym1,texturesym2,texturesym3,texturesym4,
texturesym5,texturesym6,texturesym7,texturesym8,texturesym9} in 
the context of zero textures in the three-neutrino picture. Since the sterile 
neutrino is a singlet under the standard model gauge symmetry, 
one can not prevent a bare mass term of Majorana type as well as a Dirac mass 
term involving the active neutrinos and the Higgs field. 
For a $4\times4$ neutrino mass matrix at eV scale, we should be able to keep 
both the Majorana and the Dirac mass term involving the sterile neutrino 
at the eV scale, 
which is unnatural unless some additional symmetries can ensure the smallness of 
these mass terms. 
This gives rise to another challenge in addition to generating the active 
neutrino mass matrix at sub-eV from the popular seesaw mechanism. 
Generating a $4\times 4$ light neutrino mass matrix within different seesaw 
frameworks have led to several studies in 
recent times~\cite{sterileearlier1, sterileearlier2, sterileearlier3, 
sterileearlier4, sterileearlier5, sterileearlier6, 
sterileearlier7, sterileearlier8, sterileearlier9, sterileearlier10, 
Borah:2016xkc, Borah:2016lrl, sterileearlier12}. 
Here we consider a simple extension of the idea proposed 
in Refs.~\cite{sterileearlier2, sterileearlier3} in order to accommodate the texture 
zero criteria or five-zeros in the $3\times3$ active block of the light neutrino 
mass matrix. 
Instead of giving an effective model based on higher dimensional operators and 
discrete symmetries as was done in order to 
explain the four-zero texture mass matrix in Ref.~\cite{Borah:2016xkc}, here we give a 
renormalizable model.

The particle content of the proposed model is shown in Table \ref{tab3}. 
We are showing only the fields responsible for neutrino mass generation here 
skipping the details of 
quarks and the charged lepton sector. The gauge symmetry of the standard model is 
extended by another gauge symmetry $U(1)_{L_{\mu}-L_{\tau}}$~\cite{LmLt,LmLt_1,LmLt_2}. 
Interestingly, the requirement of anomaly cancellation in a model with 
$U(1)_{L_{\mu}-L_{\tau}}$ gauge symmetry does not require any other 
fermion content apart from the usual standard model ones. Three additional 
fermions namely, $\nu_s, N_{\mu}, N_{\tau}$ are added with such 
choices of $U(1)_{L_{\mu}-L_{\tau}}$ charges that do not introduce any 
anomalies. 
Out of these three fermions, $\nu_s$ is the light sterile neutrino of eV scale 
while the other two are heavy neutrinos. 
The scalar sector of the model also consists of three additional scalar fields 
$H_2, \chi_1, \chi_2$ apart from the standard model 
Higgs field $H_1$. There also exists an approximate global symmetry $U(1)_S$ 
required to keep the bare mass term of sterile 
neutrino $\nu_s$ absent from the Lagrangian. The Yukawa Lagrangian involving the 
leptonic fields can be written as

\begin{widetext}
\begin{align}
\mathcal{L}_{\text{Yukawa}} & \supset \frac{1}{2}(Y^{\prime}_e \bar{L}_e H_1 
e_R 
+Y^{\prime}_{\mu} \bar{L}_{\mu} H_1 \mu_R+Y^{\prime}_{\tau} \bar{L}_{\tau} H_1 
\tau_R)+ Y_{\mu} \bar{L}_{\mu} \widetilde{H}_1 N_{\mu}+Y_{\tau} \bar{L}_{\tau} 
\widetilde{H}_1 N_{\tau}   \nonumber \\
& +M_N N_{\mu} N_{\tau}+ Y_{s{\mu}} \nu_s N_{\mu} \chi_1 + Y_{s{\tau}} \nu_s N_{\tau} 
\chi_2+Y_{s} \bar{L}_{e} \widetilde{H}_2 \nu_s +\text{h.c.}
\end{align}
\end{widetext}
The relevant part of the scalar potential can be written as
\begin{widetext}
\begin{align}
\mathcal{L}_{\text{Scalar}} & \supset -\mu^2_{11} H^{\dagger}_1 H_1 + \mu^2_{22} H^{\dagger}_2 H_2 +\lambda_1 (H^{\dagger}_1 H_1)^2+\lambda_2 (H^{\dagger}_2 H_2)^2 + \lambda_3 (H^{\dagger}_1 H_1) (H^{\dagger}_2 H_2) + \lambda_4 (H^{\dagger}_1 H_2) (H^{\dagger}_2 H_1) \nonumber \\
& - (\mu^2_{12} (H^{\dagger}_1 H_2)  + \text{h.c.}) - \mu^2_1 \chi^{\dagger}_1 \chi_1 + \lambda_{5} (\chi^{\dagger}_1 \chi_1)^2  - \mu^2_2 \chi^{\dagger}_2 \chi_2 + \lambda_{6} (\chi^{\dagger}_2 \chi_2)^2
\end{align}
\end{widetext}

Denoting the vacuum expectation values (vev) of the neutral components of the 
scalar fields as 
$ \langle H^0_1 \rangle =v_1, \langle \chi_1 \rangle =u_1, \langle \chi_2 
\rangle =u_2$, we can derive the leptonic mass matrices. 
The charged lepton mass matrix is diagonal and takes the form $M_L = \frac{1}{2}\text{diag} 
( Y^{\prime}_e v_1, Y^{\prime}_{\mu} v_1, Y^{\prime}_{\tau} v_1)$. 
The neutral fermion mass matrix in the basis $(\nu_L, N, \nu_s)$ can be written 
as
\begin{equation}
\mathcal{M}= \left( \begin{array}{ccc}
              0 & M_{D} &  0 \\
              M^T_{D} & M_{R} & M^T_S \\
              0 & M_S & 0
                      \end{array} \right)
\label{eqn:numatrix}       
\end{equation}
Here $M_D$ is the $3\times2$ Dirac neutrino mass matrix written in $(\nu_L, N)$ 
basis as
\begin{equation}
M_D= \left( \begin{array}{cc}
              0 & 0 \\
              Y_{\mu}v_1 & 0 \\
              0 & Y_{\tau}v_1
                      \end{array} \right)
\label{eqn:MD}       
\end{equation}
The other two matrices are given as 
\begin{equation}
M_R= \left( \begin{array}{cc}
              0 & M_N \\
              M_N & 0 
                      \end{array} \right), \;\; M_S= \left( \begin{array}{cc}
              Y_{s{\mu}}u_1 & Y_{s{\tau}} u_2
         \end{array} \right)
\label{eqn:MRS}       
\end{equation}
In the case where $M_R \gg M_S > M_D$, the effective $4\times4$ light neutrino 
mass matrix in the basis $(\nu_L, \nu_s)$ can be written as~\cite{sterileearlier2}
\begin{equation}
M_{\nu}= -\left( \begin{array}{cc}
              M_{D} M^{-1}_R M^T_D &  M_D M^{-1}_R M^T_S\\
              M_S (M^{-1}_R)^TM^T_{D} & M_SM^{-1}_{R}M^T_S
                      \end{array} \right)
\label{eqn:numatrix2}       
\end{equation}
Using the above definitions of $M_D, M_R, M_S$, the light neutrino mass matrix 
is 
\begin{equation}
M_{\nu}= -\left( \begin{array}{cccc}
              0 & 0 & 0 &  0\\
              0 & 0 & Y_{\mu} Y_{\tau} \frac{v^2_1}{M_N} & Y_{\mu} Y_{s{\tau}} 
\frac{u_2v_1}{M_N} \\
              0 & Y_{\mu} Y_{\tau} \frac{v^2_1}{M_N} & 0 & Y_{\tau} Y_{s{\mu}} 
\frac{u_1v_1}{M_N}\\
              0 & Y_{\mu} Y_{s{\tau}} \frac{u_2v_1}{M_N} & Y_{\tau} Y_{s{\mu}} 
\frac{u_1v_1}{M_N} & 2  Y_{s{\mu}} Y_{s{\tau}} \frac{u_1 u_2}{M_N}
                      \end{array} \right)
\label{eqn:numatrix3}       
\end{equation}
The second Higgs doublet $H_2$ is assumed to have a positive mass squared term, 
preventing it from acquiring a vev. However, 
after electroweak symmetry breaking (EWSB), it can acquire an induced vev due to 
the existence of terms like $\mu^2_{12} H^{\dagger}_1 H_2$ in the Lagrangian. 
This term also breaks the $U(1)_S$ global symmetry explicitly and hence prevents 
the formation of massless Goldstone boson due to 
the spontaneous breaking of continuous global symmetry. By naturalness argument, 
one can also take this soft $U(1)_S$ breaking mass term $\mu^2_{12}$ to be 
small. 
The induced vev will be $\langle H^0_2 \rangle=v_2 \approx 
\frac{\mu^2_{12}}{M^2_2}v_1$ where $M^2_2$ is given by
$$ M^2_2 = \mu^2_{22}+ \lambda_3 v^2_1 + \lambda_4 v^2_1$$
This mechanism was also adopted earlier within the three light neutrino scenarios. For example, one may refer to the work \cite{Davidson:2009ha} and references therein. Thus, one can tune the soft-breaking mass term in order to generate a small vev 
$v_2$. 
For example, if $\mu_{12} \sim 100$ keV, then for electroweak scale $\mu_2$, the 
vev is $v_2 \approx 10^{-10} \; \text{GeV} \sim 0.1 \; \text{eV}$. 
Such a tiny vev can generate a non-zero $(14)$ term of the light neutrino mass 
matrix for $\mathcal{O}(1)$ Yukawa coupling. Therefore, the final light neutrino mass 
matrix is 
\begin{equation}
M_{\nu}= -\left( \begin{array}{cccc}
              0 & 0 & 0 &  Y_{s} v_2\\
              0 & 0 & Y_{\mu} Y_{\tau} \frac{v^2_1}{M_N} & Y_{\mu} Y_{s{\tau}} 
\frac{u_2v_1}{M_N} \\
              0 & Y_{\mu} Y_{\tau} \frac{v^2_1}{M_N} & 0 & Y_{\tau} Y_{s{\mu}} 
\frac{u_1v_1}{M_N}\\
              Y_{s} v_2 & Y_{\mu} Y_{s{\tau}} \frac{u_2v_1}{M_N} & Y_{\tau} 
Y_{s{\mu}} \frac{u_1v_1}{M_N} & 2  Y_{s{\mu}} Y_{s{\tau}} \frac{u_1 u_2}{M_N}
                      \end{array} \right)
\label{eqn:numatrix4}       
\end{equation}
which resembles the structure of the five-zero texture mass matrix which is 
found to be allowed by the present data, in our analysis. 
The additional gauge symmetry of the model that is, $U(1)_{L_{\mu}-L_{\tau}}$ 
will be broken by the vev's of $\chi_{1,2}$ resulting in a massive neutral gauge 
boson $Z_{\mu \tau}$.

\section{Possible Implications}
\label{sec5b}

\begin{figure*}
\begin{center}
\includegraphics[width=0.4\textwidth]{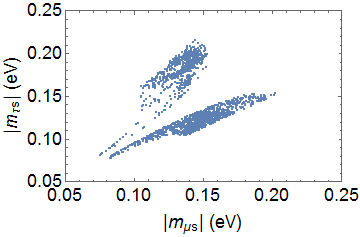}
\includegraphics[width=0.4\textwidth]{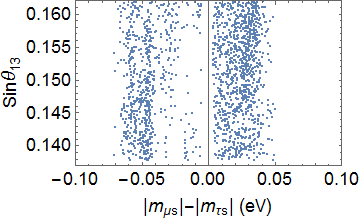} \\
\end{center}
\begin{center}
\caption{Breaking of $\mu-\tau$ symmetry in the active-sterile sector and 
generation of non-zero $\theta_{13}$.}
\label{fig2}
\end{center}
\end{figure*}

\begin{figure*}
\begin{center}
\includegraphics[width=0.45\textwidth]{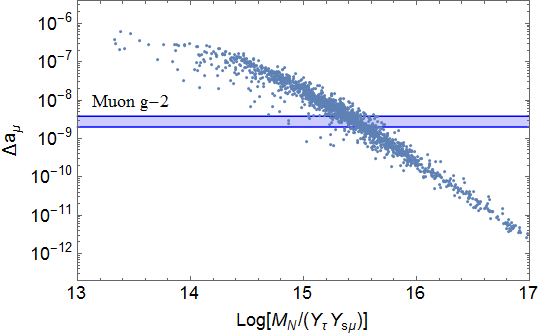}
\includegraphics[width=0.45\textwidth]{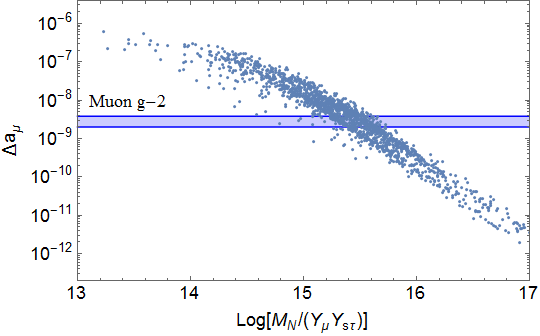} \\
\end{center}
\begin{center}
\caption{Correlation plots between muon $(g-2)$ and the model parameters.}
\label{fig3}
\end{center}
\end{figure*}

Since the zero texture models predict specific values of neutrino parameters, 
they can have very interesting implications in the lepton flavour sector. 
Here we discuss two such possible implications of the five-zero texture model discussed 
above. The first implication is the flavour symmetric origin of the five-zero 
texture mass matrix. 
We have shown that the only allowed five-zero texture mass matrix in the $3+1$ 
scenario is the one having the following structure 
\begin{equation}
m_{\nu} = \begin{pmatrix}
0 & 0 &0 & m_{es}\\
0 & 0 &m_{\mu \tau} & m_{\mu s} \\
0 & m_{\mu \tau} & 0 &m_{\tau s} \\
m_{es} &m_{\mu s} & m_{\tau s} & m_{s s}
\end{pmatrix}
\end{equation}
which has been denoted as texture $B$ in the above discussion. The $3\times3$ 
active neutrino block of this mass matrix is symmetric with respect to $\mu 
\leftrightarrow \tau$. 
Such $\mu-\tau$ symmetric $3\times3$ light neutrino mass matrix can be realised 
naturally within discrete flavour symmetry 
models~\cite{Shimizu:2011xg,Ishimori:2010au,Grimus:2011fk,King:2013eh}. 
In fact, prior to the discovery of non-zero reactor mixing angle $\theta_{13}$, 
$\mu-\tau$ symmetric light neutrino mass matrices were consistent with 
experimental data. 
This class of models predicts $\theta_{13} = 0$ and $\theta_{23} = \frac{\pi}{4}$ 
whereas the value of $\theta_{12}$ depends upon the particular model. 
Since the latest neutrino oscillation data is not consistent with 
$\theta_{13}=0$, one has to go beyond the minimal $\mu-\tau$ symmetric 
framework. 
It is interesting to note that the five-zero texture model discussed in this 
work has a $\mu-\tau$ symmetric active neutrino block and is still consistent 
with the latest neutrino oscillation data. 
This is possible due to the breaking of $\mu-\tau$ symmetry in the 
active-sterile sector $m_{\mu s} \neq m_{\tau s}$. 
This interesting possibility of generating non-zero $\theta_{13}$ in $3+1$ light 
neutrino framework has been explored in Refs.~\cite{sterilemutau0, sterileearlier2, 
sterileearlier4, sterilemutau1, sterilemutau2, sterilemutau3} and within $A_4$ 
flavour symmetric model recently in Ref.~\cite{sterilemutau4}. Although we 
are not discussing such discrete flavour symmetry in this work, the $\mu-\tau$ 
symmetric active neutrino block of the light neutrino mass matrix could be 
hinting at such a symmetry at the fundamental level. The breaking of $\mu-\tau$ 
symmetry in the active-sterile block can be seen by plotting the respective mass 
matrix elements $m_{\mu s}, m_{\tau s}$ for those values of neutrino parameters 
which satisfy the texture zero conditions of the five-zero texture mass matrix. 
From the top left panel of Fig.~\ref{fig2}, it is clear that for all the 
allowed points $\lvert m_{\mu s} \rvert \neq \lvert m_{\tau s} \rvert$. This is 
also clear from the top right panel showing $\sin{\theta_{13}}$ 
versus $\lvert m_{\mu s} \rvert -\lvert m_{\tau s} \rvert$ where none of the 
$\sin{\theta_{13}} \neq 0$ values correspond to $\lvert m_{\mu s} \rvert = 
\lvert m_{\tau s} \rvert$. The deviation from the exact $\mu-\tau$ symmetry in 
the $4 \times 4$ light neutrino mass matrix also generates deviations from 
maximal atmospheric mixing angle $\theta_{23} = \frac{\pi}{4}$. This is visible 
from the bottom right panel of Fig.~\ref{fig1} where all the allowed points 
correspond to non-maximal values of the atmospheric mixing angle.

Another interesting implication the model can have is related to the discrepancy 
in the anomalous magnetic moment of muon $(g-2)$ from the Standard Model 
prediction~\cite{muong2}. 
This discrepancy between the experimentally observed and the predicted value of 
muon $(g-2)$ is 
\begin{equation}
\Delta a_{\mu} = a^{\text{exp}}_{\mu}-a^{\text{pred}}_{\mu} = (29.0 \pm 9.0) \times 
10^{-10}
\end{equation}
The $U(1)_{L_{\mu}-L_{\tau}}$ gauge symmetric extension of the Standard Model 
discussed above can give rise to a one-loop contribution to muon $(g-2)$ with 
the $Z_{\mu \tau}$ gauge boson in the loop. 
The contribution of this one-loop diagram to muon $(g-2)$ is given 
by~\cite{Gninenko:2001hx,Baek:2001kca}
\begin{equation}
\Delta a_{\mu} = \frac{g^2_{\mu \tau}}{8 \pi^2} \int^1_0 dx \frac{2x 
(1-x)^2}{(1-x)^2+rx}
\end{equation}
where $r=(M_{Z_{\mu \tau}}/m_{\mu})^2$ and $g_{\mu \tau}$ is the 
$U(1)_{L_{\mu}-L_{\tau}}$ gauge coupling. 
As the $U(1)_{L_{\mu}-L_{\tau}}$ gauge symmetry is spontaneously broken by the 
vev's of $\chi_{1,2}$, the corresponding gauge 
boson mass can be written as $M_{Z_{\mu \tau}}=g_{\mu \tau} \sqrt{u^2_1+u^2_2} = 
\sqrt{2} g_{\mu \tau} u$ assuming $u_1=u_2$. 
These vev's also appear in the light neutrino mass matrix through the 
active-sterile sector. For example, the $m_{\mu s}, m_{\tau s}$ elements are 
given as 
$$m_{\mu s} =Y_{\mu} Y_{s{\tau}} \frac{uv_1}{M_N}, m_{\tau s}=Y_{\tau} 
Y_{s{\mu}} \frac{uv_1}{M_N}$$
Since the vev $u$ appear in the expression for gauge boson mass and hence in 
$\Delta a$, we can relate the model parameters $M_N, Y_{\mu, \tau}, Y_{s \mu, 
s\tau}$ to muon $(g-2)$ as the 
light neutrino mass matrix elements $m_{\mu s}, m_{\tau s}$ are predicted by the 
texture zero conditions. 

It should be noted that there are several constraints on the mass and coupling of the extra neutral gauge boson $Z_{\mu \tau}$. If the gauge boson is light $M_{Z_{\mu \tau}} \leq 1 \; \text{MeV}$, it can open new decay channels of mesons $ K^-, \pi^- \rightarrow \mu^- \overline{\nu_{\mu}} Z_{\mu \tau}$. The experimental limits on such new decay channels constrain the corresponding gauge coupling $g_{\mu \tau} \leq 10^{-2}$ \cite{Lessa:2007up}. Another constraint on $(g_{\mu \tau}, M_{Z_{\mu \tau}})$ comes from the experimental measurement of neutrino trident processes like $\nu_{\mu} N \rightarrow \nu_{\mu} N \mu^+ \mu^-$ where $N$ denotes a nucleus. The CCFR measurement of the neutrino trident cross-section rules out a part of the parameter space in the $(g_{\mu \tau}, M_{Z_{\mu \tau}})$ plane \cite{Altmannshofer:2014pba}. In the low mass regime $1\; \text{MeV} < M_{Z_{\mu \tau}} < 1\; \text{GeV}$ which is of our interest, this experimental bound corresponds to approximately $g_{\mu \tau} \leq 8 \times 10^{-4}$. Though this upper bound slightly gets relaxed as $M_{Z_{\mu \tau}}$ is increased from 1 MeV to 1 GeV, we consider the most conservative bound in our analysis. On the other hand, cosmology can also constrain a light gauge boson $Z_{\mu \tau}$ which couple to the light neutrinos. The Planck bound on the number of effective relativistic degrees of freedom \cite{Ade:2015xua} during the epoch of BBN constrains the mass of such additional gauge bosons coupling to neutrinos as $M_{Z_{\mu \tau}} \geq 5 \; \text{MeV}$ \cite{Kamada:2015era}. Therefore, all these constraints can be simultaneously taken into account if we consider $g_{\mu \tau} \leq 8 \times 10^{-4}, M_{Z_{\mu \tau}} \geq 5 \; \text{MeV}$. Hence, we vary them in the range $ 10^{-5} \leq g_{\mu \tau} \leq 8 \times 10^{-4}, 5 \; \text{MeV} \leq M_{Z_{\mu \tau}} \leq 1 \; \text{GeV}$ for the purpose of our numerical analysis discussed below.

We show the variation of $\Delta a_{\mu}$ with these 
parameters in Fig.~\ref{fig3}. In this plot, we consider a light gauge boson mass $M_{Z_{\mu \tau}}$ in order to have maximum effect on muon $(g-2)$ through one-loop effects. 
This is possible even if the $U(1)_{L_{\mu}-L_{\tau}}$ gauge symmetry is broken at a high scale $u_{1,2} \geq \mathcal{O} (\text{TeV})$ due to tiny 
gauge coupling $g_{\mu \tau}$ which appears in gauge boson mass expression mentioned above. 
To be more specific, we randomly vary the gauge coupling and gauge boson mass in the range $g_{\mu \tau} \in (10^{-5}, 8\times10^{-4}), M_{Z_{\mu \tau}} \in (5\times10^{-3}, 1) \; \text{GeV}$ and calculate 
the predictions for $\Delta a_{\mu}$ and show it as a function of neutrino mass matrix parameters $M_N, Y_{\mu, \tau}, Y_{s \mu, 
s\tau}$ in Fig.~\ref{fig3}. It is interesting to see from these panels that the 
requirement of explaining the muon $(g-2)$ anomaly restricts the ratio of heavy 
neutrino mass to the product of Yukawa to a very narrow range.

\section{Conclusion}
\label{sec6}

In this paper we have discussed the viability of various zero texture conditions 
in low energy Majorana neutrino mass matrix in the 3+1 scheme. Each 
element in the neutrino mass matrix is a functions of neutrino masses, 
mixing angles, Dirac phases 
and Majorana phases. As it is not possible to measure all of them directly in 
the experiments, the zero texture conditions are proposed to reduce the 
parameter space and to obtain 
various correlations among different parameters which can be 
verified/falsified in different experiments. The results of zero textures 
in $m_{\nu}$ are also important for testing various neutrino mass
models. In 3+1 scheme, the number of independent elements are 10 and thus in 
principle it is possible to have minimum one-zero texture 
(when one of the elements in the neutrino mass matrix is zero) to maximum ten 
zero textures. Earlier studies show that a zero texture involving 
the sterile elements (i.e.,
the elements belonging to the fourth row/column) is not possible. This leaves us 
with at most six-zero textures. Earlier studies have also 
explored
the possibility of one, two, three and four-zero textures in the 3+1 scheme. In the 
present work we have discussed the main findings of these past studies and also 
studied the
remaining zero texture conditions i.e., five and six-zero textures. While 
discussing the past results, 
our main observation is that the results of the zero textures in 3+1 scheme, 
heavily depends upon the choice of sterile mixing parameters. 
%\st{and the method of 
%solving the nonlinear algebraic complex
%equations}. 
In our original analysis of five-zero textures we find that among the 
six possible structures only one is allowed by the current oscillation data in 
normal hierarchy. 
We have also presented the prediction of $\theta_{14}$ and the Majorana phase 
$\gamma$ for this allowed texture.
We also showed that the viability of this texture demands: (i) 
$\theta_{14}$/$\theta_{24} < 3^\circ$, (ii) $3.4^\circ < \theta_{34} < 11^\circ$ 
and (iii) $\theta_{24} \neq 45^\circ$. 
We have outlined one possible symmetry realisation of this five-zero 
texture mass matrix by incorporating an anomaly free $U(1)_{L_{\mu}-L_{\tau}}$ 
gauge symmetry. Such a gauge symmetry can not only explain the structure of the 
five-zero texture mass matrix, but can also give rise to other observable 
consequences. We discuss one such possibility in terms of the anomalous magnetic 
moment of the muon and show that the model can explain the anomaly with 
reasonable values of different couplings. We also briefly discuss the discrete 
$\mu-\tau$ symmetry possessed by the $3\times 3$ active block of the light 
neutrino mass matrix and the related implications. To summarise, from the 
results obtained in this work we understand that the five-zero textures are the 
maximum allowed textures in 3+1 scheme and more than five zero textures in the low energy
neutrino mass matrix are not allowed. We believe our present work is a 
comprehensive analysis of zero textures in 3+1 scheme and the results discussed 
here will be important to testify the existence of sterile neutrino and also for 
building models for light sterile neutrinos.

\section*{Acknowledgements}

We thank Suprabh Prakash for useful discussions. 
The work of MG is partly supported by the ``Grant-in-Aid for Scientific Research 
of the Ministry of Education, 
Science and Culture, Japan", under Grant No. 25105009. The work of SG is 
supported by the Australian Research 
Council through the ARC Center of Excellence in Particle Physics (CoEPP 
Adelaide) at the Terascale (CE110001004). SKR acknowledges support from 
IBS under the project code IBS-R018-D1.

%%%%%%%%%%%%%%%%%%%%%%%%%%%%%%%%%%%%%%%%%%%%%%%%%%%%%%%%%%%%%%%%%%%%%%%%%%%%%%%%
%%%%%

\end{document}